\documentclass[superscriptaddress,pra,twocolumn,aps,showpacs]{revtex4}
\usepackage{epsfig}
\usepackage{bm}
\begin{document}
\date{\today}

\title{Shell-Model Monte Carlo Simulations of BCS-BEC Crossover in Few-Fermion Systems}
\author{N.T. Zinner}\email{zinner@phys.au.dk}
\affiliation{Department of Physics, Harvard University,
Cambridge, Massachusetts 02138, USA}
\affiliation{Department of Physics
and Astronomy, Lundbeck Foundation Theoretical Center for Quantum Systems Research, University of Aarhus,
DK-8000 {\AA}rhus C, Denmark}
\author{K. M\o lmer}
\affiliation{Department of Physics
and Astronomy, Lundbeck Foundation Theoretical Center for Quantum Systems Research, University of Aarhus,
DK-8000 {\AA}rhus C, Denmark}
\author{C. \"Ozen}
\affiliation{Department of Physics, Yale University, New Haven, Connecticut 06520, USA}
\affiliation{GSI Helmholtzzentrum f\"ur Schwerionenforschung, D-64259 Darmstadt, Germany}
\author{D.J. Dean}
\affiliation{Physics Division, Oak Ridge National Laboratories, Oak Ridge, Tennessee 37831, USA}
\author{K. Langanke}
\affiliation{GSI Helmholtzzentrum f\"ur Schwerionenforschung, D-64259 Darmstadt, Germany}
\affiliation{Technische Universit\"at Darmstadt, D-64289 Darmstadt, Germany}
\affiliation{Frankfurt Institute for Advanced Studies, D-60438 Frankfurt, Germany}

\begin{abstract}
We study a trapped system of fermions with a zero-range two-body
interaction using the shell-model Monte
Carlo method, providing {\em ab initio} results for the low
particle number limit where mean-field theory is not applicable. We present results for
the $N$-body energies as function of
interaction strength, particle number, and temperature. 
The subtle question of renormalization
in a finite model space is addressed and the convergence of our method and its
applicability across the BCS-BEC crossover is discussed. Our findings indicate that
very good quantitative results can be obtained on the BCS side, whereas at unitarity
and in the BEC regime the convergence is less clear.
Comparison to $N=2$ analytics at zero and finite temperature, and to other 
calculations in the literature for $N>2$ show very good agreement.
\end{abstract}

\pacs{03.75.Ss, 05.30.Fk, 21.60.Ka}
\maketitle

\section{Introduction}
The physics of ultracold atomic gases has been intensively pursued
experimentally and theoretically in the last decade. Recently there
has been great interest in strongly-interacting Fermi
gases where Feshbach resonances allow the tuning of the two-body interaction, and
studies of the transition
from a dilute gas of fermionic atoms to a Bose condensate of
molecules are now possible in the laboratory
\cite{ohara2002,dieckmann2002,chin2004,zwierlein2005,kinast2005,stewart2006}.
While studies of degenerate Fermi gases have mostly dealt with large
atom numbers and wide traps, efforts have begun to trap only a few
atoms (1-100) in tighter traps \cite{jochim2008}. Also, with the
implementation of three-dimensional optical lattices, a
low-tunneling regime can be reached with essentially
isolated harmonic oscillators containing only a few fermions at
each site \cite{stoferle2006}. This means that one can now
explore few-body fermionic effects in trapped systems with
scattering lengths that are comparable to the inter-particle distance
and the trap width.

In this paper we report on a theoretical study of harmonically
trapped fermions using the shell-model Monte Carlo (SMMC) approach.
This method has been extensively used in nuclear physics to
determine nuclear properties at finite temperature in larger model
spaces than can be handled by normal nuclear shell-model
diagonalization \cite{johnson1992,koonin1997}. In the SMMC, the
many-body problem is described by a canonical ensemble at
temperature $k_B T=\beta^{-1}$ and the Hubbard-Stratonovich
transformation is used to linearize the imaginary-time many-body
propagator $e^{-\beta H}$. Observables are then expressed as path
integrals of one-body propagators in fluctuating auxiliary fields.
The method is in principle exact and subject only to statistical
uncertainties.
For equal mixtures of two hyperfine
states at low density, the interaction can be modeled
with an $s$-wave zero-range potential. Importantly, this interaction
is free of sign problems \cite{cem2009} that are otherwise known to plague 
quantum Monte Carlo simulations with fermions. 
We present here the first application of this many-body
method to ultracold gas physics.

Previous works have considered few-fermion systems using advanced many-body methods. The Green's function
Monte Carlo
methods were applied to homogeneous \cite{carlson2003}, as well as trapped systems \cite{bertsch2007}.
No-core \cite{stetcu2007}, and traditional shell-models \cite{alhassid2007}, using effective interactions
have also recently been applied to these systems, particularly for very low particle numbers where exact
results are available \cite{werner2006a,werner2006b}. Finite-temperature, non-perturbative lattice methods
have
also been applied to the homogeneous case \cite{bulgac2007,akkineni2007}.
These works mostly focus on 
the unitary $|a|\rightarrow \infty$ limit and the crossover regime around it.
Most previous Monte Carlo approaches have used fixed nodes in the many-body wave function 
in order to alleviate the sign problem, making the methods variational.
As we will now demonstrate the present method has no sign problem \cite{cem2009} and 
can be used on the BCS side, in the crossover
region, and also into the BEC regime.

\section{Model and Renormalization Scheme}
The model Hamiltonian used is
\begin{equation}
H=\sum_i \frac{p_{i}^{2}}{2m}+\sum_i \frac{1}{2}m\omega^2 r_{i}^{2}+\sum_{[ij]}V_0
\delta(\vec{r}_i-\vec{r}_j),
\end{equation}
where we sum over all particles $i$ and $[ij]$ denotes a sum over
fermion pairs with opposite internal (hyperfine) states.
The trap frequency is
$\omega$, and $V_0$ denotes the interaction strength.
The SMMC method was originally set up to handle nucleons where the Hamiltonian above appears 
in extensively studied pairing
problems in nuclei \cite{dean2003}. The two-component
Fermi gas can now be mapped onto a single spin $1/2$ nucleon species \cite{heiselberg03,cem2009}.

Dimensional arguments reveal that the matrix elements needed in a shell-model 
approach scale as
$1/b^3$, where $b=\sqrt{\hbar/m\omega}$ is the oscillator length. It is therefore natural
to redefine the interaction strength in terms of $V_0=-g \hbar\omega
b^3$, where $g$ is a dimensionless strength measure. In order to relate to physical quantities
the strength of the zero-range interaction must be regularized \cite{weinberg1991}. 
The shell-model works in finite model spaces and this naturally introduces a cut-off in energy 
$E_c=\alpha^2\hbar\omega$, where $\alpha^2=N_{max}+3/2$. 
Regularization in finite model spaces with discrete energies is notoriously difficult and several
prescriptions have been adopted in the literature. Here we will use the simplest 
strategy and renormalize the coupling $g$ through low-energy scattering parameters 
defined in the continuum \cite{esbensen1997}. This defines a relation between $g$, the $s$-wave 
scattering length $a$, and $E_c$, 
which is $4\pi a/b=g/(\alpha g/2\pi^2-1)$. This yields the effective interaction
strength for a model space with $N_{max}+1$ major harmonic oscillator shells and gives a prescription
for varying the interaction strength with model space size in order to keep $a$ fixed.

The relevant parameter regime is
expected to be where the natural energy scale, given by the level spacing $\hbar\omega$, 
is comparable to typical
two-body matrix elements between the trap states at $T=0$ (as a quantitative measure we use
$\langle 1s1s|V|1s1s\rangle$). This turns out to be around $g\sim 10$ in
our setup at $N_{max}=3$, which corresponds to $a=11 b$ in the continuum regularization scheme. The
regions of interest in terms of interaction strength and
$\hbar\omega$ for atomic gases
are discussed in \cite{heiselberg2002,bruun2002}.

\begin{figure}
\begin{center}
\epsfig{file=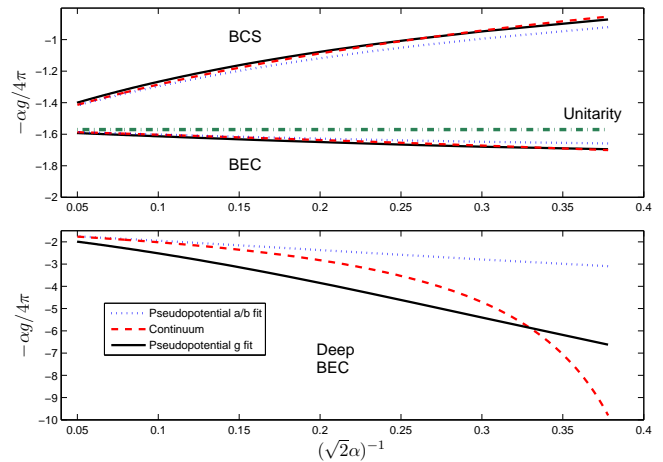,clip=true,scale=0.43}
\caption{(Color online) Running of the coupling, $-\alpha g/4\pi$, in continuum and pseudopotential regularization schemes as function of the inverse cut-off $(\sqrt{2}\alpha)^{-1}$. The upper panel shows
the $a/b=-1.0$ (BCS) and $a/b=11$ (BEC) cases used in the calculations. The dashed line
is the continuum scheme with the couplings $g=10$ (BEC) and $g=5.35$ (BCS) at $N_{max}=3$, whereas the solid line is the pseudopotential 
line that runs through the same couplings at $N_{max}=3$. The dotted line is the pseudopotential coupling for the same $a/b$ ratio as in the continuum case. The dot-dashed line is unitarity $|a|=\infty$. 
The lower panel demonstrates the deviations in the deep BEC
regime $a/b=1.0$. Notice the difference in scales on the vertical axis between the upper and lower panels.}
\label{fig1}
\end{center}
\end{figure}

It is appropriate to discuss alternative renormalization schemes that were recently 
proposed \cite{alhassid2007,stetcu2007} which take the trap level spacing into account in a direct
way inspired by effective field theory techniques. The idea is to relate the bare coupling $V_0$ to $a$ by calculating the two-body ground state energy
in the finite model space of harmonic oscillator states with energy less than $E_c$ and then determining
$a$ through the exactly solvable pseudopotential model of Busch {\it et al.} \cite{busch1998}, 
i.e. solving the equation $\Gamma(3/4-E/2\omega)/\Gamma(1/4-E/2\omega)=b/\sqrt{2}a$, with $\Gamma$ the Gamma function.
There has been some experimental support of the Busch {\it et al.} results
for fermion pairs in three-dimensional (3D) optical lattices \cite{stoferle2006}. However, the pseudopotential is only expected 
to be correct when the van der Waals scale is much smaller than the trap, and 
since we have tight traps with only few fermions in mind this is not necessarily the case. We therefore expect
both the continuum and the pseudopotential approaches to be approximations for a tight trap \cite{bolda2002}.
In Fig.~\ref{fig1} we have plotted the running of the coupling in both schemes similarly to Fig.~1 of \cite{stetcu2007}.
The upper part shows the two cases used in our calculations, whereas the lower part is in the deep BEC regime ($a/b=1.0$).
The two approaches yield $a/b=11$ (continuum) and $a/b=7.8$ (pseudopotential) for
$g=10$ with $N_{max}=3$, which is not too severe considering the divergence of $a$ at unitarity.
In the case
of $g=5.35$ at $N_{max}=3$ we have $a/b=-1.0$ (continuum) and $a/b=-0.9$ (pseudopotential).
In the exact model of
\cite{busch1998} this translates to a difference in ground state energy of only $0.03\hbar\omega$ for both cases.
The running of the coupling is the same to 
two significant digits for both $a/b=11$ and $a/b=-1.0$ results for $N_{max}=2,$ 3, and 4 used here. If we turn the 
argument around and instead fix the ratio $a/b$, the pseudopotential couplings are given as the dashed blue curves
in Fig.~\ref{fig1}. For the $a/b=-1.0$ case, this gives $g=5.67$ as compared to our $g=5.35$, and $g=9.79$ compared to
$g=10$ for $a/b=11$. These modifications would only bring us slightly downward on the BCS side and slightly upward close
to unitarity.
Only on the deeper
BEC side of the crossover ($5\gtrsim a/b>0$ in the continuum scheme), where the pairs are essentially molecules, 
do we find larger deviations of the coupling in the two schemes. The lower panel in Fig.~\ref{fig1} shows the $a/b=1.0$
case. Here one can clearly see that the running of the schemes is very different expect for very large model spaces.
We do not expect the SMMC approach to work in the molecular regime.

\section{Results}
\begin{table}
\caption{Energies (in units of $\hbar\omega$) calculated with the SMMC method for a trapped fermion gas
with scattering lengths
$a/b=11$ (BEC) and $a/b=-1.0$ (BCS) calculated at temperature $k_B T=1/5\hbar\omega$
for different particle numbers $N$. The statistical uncertainty is given in parenthesis.
HOSD denotes the non-interacting energies at $T=0$.}
\begin{tabular}{|c|ccc||c|ccc|}
\hline
N  & HOSD & BEC & BCS   &   N & HOSD & BEC & BCS  \\
\hline
2  & 3  & 1.72(3) & 2.49(3)  & 12 & 32 &20.7(2)&27.23(2)    \\
3  & 5.5 & 3.9(2)& 4.84(2)  & 13 & 35.5 & 24.0(2)&30.21(2)    \\
4  & 8    & 4.96(3)& 6.84(4)    & 14 & 39 &26.0(1)&33.16(2) \\
5  & 10.5 & 7.1(3)& 9.14(2)    & 15 & 42.5 &29.1(1)&36.08(5)  \\
6  & 13& 8.11(7)& 11.08(5)      & 16 & 46 &30.9(1)&39.03(2)   \\
7  & 15.5 &10.6(2)&13.28(5)   & 17 & 49.5 &33.6(2)&41.98(2)  \\
8  & 18 &11.58(5) &15.21(4)    & 18 & 53 &35.7(1)&44.89(2)  \\
9  & 21.5 &14.8(2)&18.30(4)   & 19 & 56.5 &39.0(1)&47.85(2)   \\
10 & 25 &16.34(6)&21.27(3)     & 20 & 60 & 40.8(1)&50.75(2)\\
11 & 28.5 & 19.3(2)&24.28(2)    & 21 & 65.4 &44.5(2)  &54.50(2)  \\
\hline
\end{tabular}
\label{tab1}
\end{table}

\subsection{Zero-temperature Limit}
\begin{figure}
\begin{center}
\epsfig{file=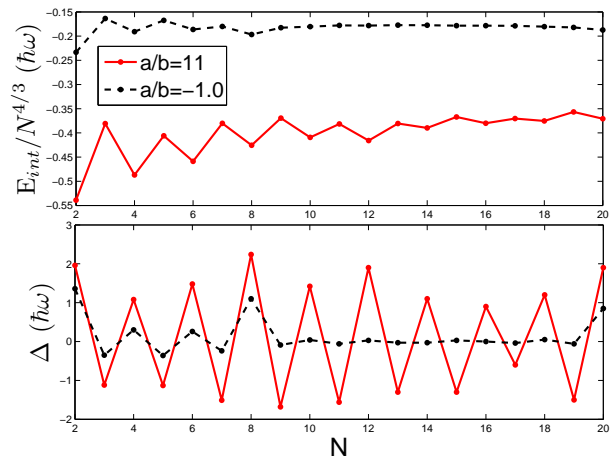,clip=true,scale=0.4}
\caption{(Color online) Scaled energies $E_{int}/N^{4/3}$, where $E_{int}=E-E_{g=0}$ (upper panel), and $\Delta(N)=E(N+1)-2E(N)+E(N-1)$ (lower panel) as a function of particle number $N$ for 
$a/b=11$ (solid) and $a/b=-1.0$ (dashed).}
\label{fig2}
\end{center}
\end{figure}

In Table \ref{tab1} we present the SMMC energies for two values
of the scattering length $a$, along with the
non-interacting energies, for particle numbers $N=2-21$. All results
in the table were calculated with a cut-off $N_{max}=3$, corresponding to a four major shell
model space. 
The results at low $T$ are well converged in this model space for $a/b=-1.0$ as we demonstrate below.
For $a/b=11$ the energies suffer larger uncertainties as discussed below.
The results
in the table were all calculated at inverse temperature $\beta=5/\hbar\omega$, above which 
we find only statistical changes in energy (see below). Thus the results in Table \ref{tab1} are
the SMMC estimates of the $T=0$ energy.
To explore the results, we plot $(E-E_{g=0})/N^{4/3}$ vs. $N$ in the upper part of Fig. \ref{fig2}, which is the
interaction energy divided by the Thomas-Fermi scaling. We can clearly see an odd-even staggering that becomes
more pronounced on the BEC side. The lower panel in Fig. \ref{fig2} shows this more clearly through the
fundamental gap $\Delta(N)=E(N+1)-2E(N)+E(N-1)$. Notice here that the closed shells are prominent and even more
so on the BCS side. We thus predict that energy measurements alone can probe the crossover through odd-even 
staggering at and away from closed shells.

As discussed above, the BEC results in Table \ref{tab1} were obtained with $g=10$ at $N_{max}=3$ or $a=11b$
which is on the BEC side of the crossover but 
still quite close to unitarity which is located at $g=9.31$ for $N_{max}=3$ (for both regularization schemes discussed above). The 
energies at unitarity are therefore slightly above our BEC results. The BCS results were obtained with $g=5.346$ or $a=-1.0b$.

As discussed, the $N=2$ problem can be solved exactly for all $a$ in a pseudopotential approach \cite{busch1998}. 
If we insert $a=11b$ we find $E=1.92\hbar\omega$
for the $T=0$ ground state, whereas for $a=-1.0b$ we find $E=2.49\hbar\omega$.
Our values of 1.72(3) and 2.49(3) at $T=1/5 \hbar\omega/k_B$ thus indicates that we are close to the $T=0$ limit. 
We notice that the energy is underestimated on the BEC side, indicating that our method
becomes worse in that regime.

\begin{figure}
\begin{center}
\epsfig{file=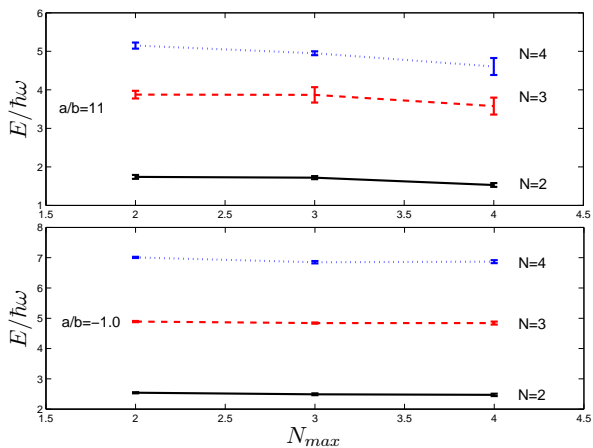,clip=true,scale=0.4}
\caption{(Color online) Energy in units of $\hbar\omega$ as function of model space size $N_{max}$ as defined in the text for 
particle numbers $N=2,$ 3, and 4 with $a/b=11$ (upper panel) and $a/b=-1.0$ (lower panel).}
\label{fig3}
\end{center}
\end{figure}

In general we find excellent agreement with previous calculations. For $N=3$ the exact 
energy at unitarity is $E=4.27\hbar\omega$ \cite{werner2006a}. The results in Table \ref{tab1} are slightly below (above) this 
value on the BEC (BCS) side, as we would expect. For higher $N$, we find good agreement with the results 
of \cite{blume2008}.
A comparison can also be made with the
results at unitarity of \cite{bertsch2007} which cover the particle numbers presented here, 
and for which our $a=11b$ energies are slightly smaller 
but sandwiched between the values in Table II of \cite{bertsch2007}. In order to elaborate on the model space size
effects we plot in Fig.~\ref{fig3} the energy as function of $N_{max}$ for $N=2,$ 3, and 4 at the converged temperature
$\beta=\hbar\omega/5$ (see below) for $a/b=11$ (upper panel) and $a/b=-1.0$ (lower panel). The convergence is
clearly much better on the BCS than on the BEC side. In particular, we see a decreasing trend of the energy with model
space size for $a/b=11$, although we remark that at least for $N=2$ and 3, 
the results are consistent within the statistical uncertainties.
We thus see that the SMMC
is robust in calculating the ground state energies of small Fermi systems on the BCS side. At unitarity and into the BEC 
regime our results are not as robust, and the deep BEC regime cannot be accessed.

\subsection{Model space size and convergence}
The SMMC method works at finite temperature and $T=0$ results are obtained by increasing $\beta=(k_BT)^{-1}$ and 
finding the point of convergence. As we work in finite model spaces and must use regularized interactions, it is 
imperative to have this convergence under control. However, as the SMMC is also a shell-model method, there is
a fundamental computational restriction on the model space size that can be used. The SMMC requires diagonalization 
of one-body terms, which grow quadratically in the number of model space states. While being more benign in its
scaling, a compromise must still be found that allows many calculations at different 
$\beta$ values with good Monte Carlo statistics. We caution that larger model spaces also lead to more numerical
noise in the calculation, and that this will grow as the coupling gets stronger.

\begin{figure}
\begin{center}
\epsfig{file=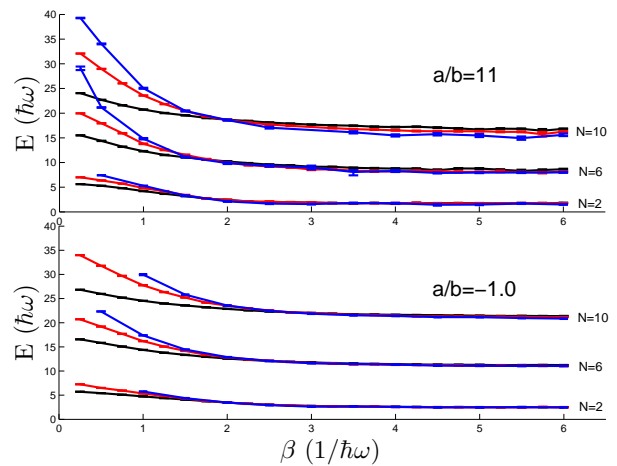,clip=true,scale=0.4}
\caption{(Color online) Energies calculated with the SMMC method 
in units of $\hbar\omega$ as a function of inverse temperature $\beta=(k_B T)^{-1}$ for $N=2,$ 6, and $10$ with strength parameter
$a/b=11$ (upper panel) and $a/b=-1.0$ (lower panel) for model space sizes $N_{max}=2$ (black), 3 (red), and 4 (blue).}
\label{fig4}
\end{center}
\end{figure}

Figure \ref{fig4} shows the SMMC energies as a function of $\beta$
for selected even particle
numbers $N=2,$ 6, and $10$ for $a=11b$ (upper panel) and $a=-1.0b$ (lower panel). The horizontal bars
indicate the error estimates from the Monte Carlo integrations. The first thing we notice is that 
the energies are virtually converged at $\beta\geq 5/\hbar\omega$ for all $N$ and
all model spaces in the $a/b=-1.0$ case. The fluctuations above this point are largely due to numerical 
noise at large $\beta$ that makes sampling difficult. This justifies our choice of $\beta=5/\hbar\omega$
in Table \ref{tab1}. We see the convergence getting worse with $N$, which is expected as there will 
be model space saturation. However, within the statistical uncertainties, we see good agreement
for the different model spaces for both $N=2$ and $6$. For $N=10$ there is a slight decrease of the energy
with model space size and our convergence is not quite as good. This is clear since for 
larger $N$ there is less space to excite particles in the given model space.
For $a/b=11$ the convergence is noticeably worse with larger fluctuations. Here the pairing is strong and a 
mixing of scales in the numerics makes the Monte Carlo sampling difficult. This problem becomes worse for larger
$g$. 
Relating back to Table \ref{tab1},
the $N\geq 10$ results are therefore probably slightly overestimated. However, the good agreement with the results
of \cite{bertsch2007} indicates that the deviations are under control and on the scale of that seen for $N=10$.

For odd $N$ the projection onto exact number states introduces a sign problem \cite{koonin1997} which grows with 
the coupling. Therefore the odd $N$ results in Table~\ref{tab1} tend to have larger uncertainties. This can also be
seen in Fig.~\ref{fig3}.
Notice that the large differences at low $\beta$ in Fig. \ref{fig4} are due to the finite model space. At 
high $T$, the particles will equilibrate in the available states and the energy becomes $E\sim N\bar{E}$, where 
$\bar{E}=\sum_i \epsilon_i/d_i$ with $\epsilon_i$ the $i$th single-particle energy and $d_i$ the corresponding 
degeneracy. This will in turn make the energy in the high-$T$ limit grow with $N_{max}$ toward its thermodynamic 
value $E(T)=3Nk_B T$ for the non-interacting harmonic oscillator.

Throughout the calculation we have used the continuum regularization method.
As noted, only on the deep BEC side will there be large differences to the pseudopotential regularization approach 
of \cite{stetcu2007,alhassid2007}. However, these renormalization schemes do not depend explicitly on $N$. In our calculations we see the convergence
becoming worse for high $N$ which might suggest an $N$ dependence; $V_0=V_0(a,b,N,N_{max})$. 
This can be improved upon by
using effective interactions in finite model spaces as done often in nuclear physics where good
effective two-body interactions do depend on $N$. We will explore such 
options in the future.

\begin{figure}
\begin{center}
\epsfig{file=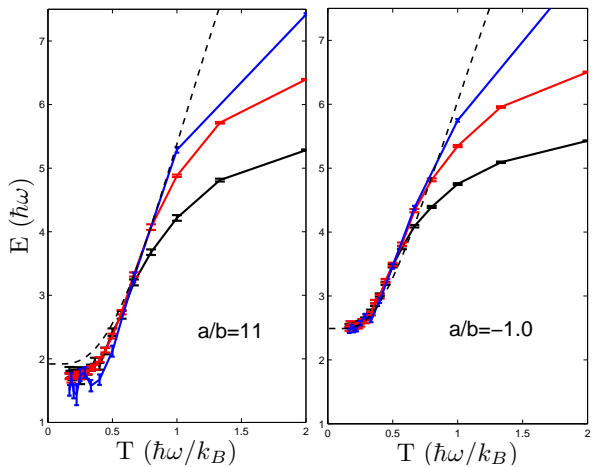,clip=true,scale=0.4}
\caption{(Color online) Energy for $N=2$ as function of temperature, $T$, in units of $\hbar\omega$ for 
$a/b=11$ (left) and $a/b=-1.0$ (right) with model spaces $N_{max}=2$ (black), $N_{max}=3$ (red), and $N_{max}=4$ (blue).
The black dashed line is the Busch {\it et al.} \cite{busch1998} result.}
\label{fig5}
\end{center}
\end{figure}

\subsection{Finite temperature $N=2$ sesults}
To explore the temperature dependence of our results, we present in Fig. \ref{fig5} the $N=2$ energies 
with $a/b=11$ (left) and $a/b=-1.0$ (right) for different model spaces. The dashed line in the figure is obtained from
the Busch {\it et al.} \cite{busch1998} results.
Here we see that the
SMMC obtains very good agreement as function of $T$ with the exact result, except for the very low and high $T$ 
regions. 
At low $T$ we see deviations and fluctuations for $a/b=11$. The latter are due to numerical problems at strong
coupling, whereas we believe the deviation from the exact result of the average 
is connected to the regularization as discussed above. 
For intermediate
$T$ we find very good agreement with the exact results, and as in Fig. \ref{fig4} 
we also see that we have obtained convergence in our
larger model spaces below $T\sim 0.8\hbar\omega/k_B$ .
The latter is
very important for finite-temperature quantities such as the specific heat, where model space sizes
can produce so-called Schottky peaks \cite{koonin1997}. For $N>2$ we have indications of a pairing
phase transition in the pair correlations and the specific heat. These results will be presented
shortly.

\section{Conclusion and Outlook}
As we have shown, the SMMC offers a good quantitative description of
the behavior of small Fermi systems in an interesting
interaction regime. We studied the case of isotropic traps and
calculated the many-body energies. The
excellent convergence properties of the method on the BCS side of the
BCS-BEC crossover
holds promise for
application to specific situations where, e.g., deformation
properties and formation of higher angular momentum pairs may become
relevant. Deformation was studied in the nuclear case using the SMMC
to predict shape transitions occurring as a function of temperature in 
the competition between pairing and quadrupole interactions \cite{koonin1997}.
This is relevant also for ultracold gases with the recent realization 
of condensates with intrinsic long-range interactions between the atoms
\cite{lahaye2007}. Recently the SMMC was also used to study the parity and 
spin properties of the density of states in nuclear systems \cite{alhassid2007b,kalmykov2007}.
Transforming this to the atomic system could help us understand the
low-energy excitation spectrum and the response of the gas to external perturbations.

\subsection*{Acknowlegments}
We thank the Helmholtz Alliance Institute EMMI for support.

\end{document}